\documentclass[preprint,nofootinbib,amsmath,amssymb,amsfonts,aps,prd,groupedaddress]{revtex4-1}

\usepackage{amssymb,amsfonts}
\usepackage{setspace} \usepackage{graphicx} \usepackage{subfigure}
\usepackage[bookmarks=true, pdfstartview={FitH}, bookmarksopen=true,
bookmarksopenlevel=1, linktoc=page]{hyperref} \usepackage{amsthm}
\usepackage{mathrsfs} 
\usepackage{tikz}
\usetikzlibrary{patterns}

\newcommand{\be}{\begin{equation}} \newcommand{\ee}{\end{equation}}
\newcommand{\ba}{\begin{equation}\begin{aligned}}
\newcommand{\ea}{\end{aligned}\end{equation}}

\DeclareMathAlphabet{\mathpzc}{OT1}{pzc}{m}{it}

\usepackage{amsthm} \let\oldproofname=\proofname
\renewcommand{\proofname}{\rm\bf{\oldproofname}:}

\usepackage{mathrsfs}

\begin{document}

\title{Jacob Bekenstein and the Development of Black Hole Thermodynamics}

\author{Robert M. Wald}
\email{rmwa@uchicago.edu}
\affiliation{Enrico Fermi Institute and Department of Physics, The University
of Chicago, 5640 South Ellis Avenue, Chicago, Illinois 60637, USA}

\begin{abstract}

I give some personal reflections on Jacob Bekenstein's pioneering work on associating an entropy to a black hole proportional to its area and on the generalized second law of thermodynamics.

\end{abstract}

\maketitle

Jacob Bekenstein and I were classmates in graduate school in Princeton. I arrived in the Autumn of 1968. Bekenstein did not arrive until 1969, but he had already completed a master's degree, so we were at about the same stage in our knowledge of physics and general relativity when we both began to do research in early 1970. Both of us were primarily interested in the theory of black holes. We had desks near each other in the newly opened Jadwin Hall, and we interacted on nearly a daily basis. We both completed and defended our Ph.D. theses in the Spring of 1972. 

General relativity was in the midst of a golden age while we were at Princeton. Penrose had proved the first singularity theorem a few years earlier, and the global methods that he thereby introduced were soon extended---particularly by him, Hawking, and Geroch---to obtain a large number of remarkable new insights into the theory. Israel's theorem on uniqueness of static black holes had very recently been proven. The remarkable global structure and other properties of the Kerr black hole had recently been fully understood, and the Penrose energy extraction process had just been proposed. Pulsars had very recently been discovered and associated with rotating neutron stars.

This golden age was reflected by the students attracted to general relativity at that time. Among the students at Princeton working on problems in general relativity with whom Bekenstein and I overlapped for at least a year were Demetrios Christodoulou, Marc Davis, Steve Fulling, Bei-Lok Hu, Bahram Mashhoon, Niall O'Murchadha, Stuart Shapiro, Claudio Teitelboim/Bunster, and Bill Unruh. I recall attending a ``town hall" meeting of faculty and graduate students sometime in 1971-72 where the particle theory faculty decried the fact that they could hardly attract any students!

The atmosphere in relativity group at Princeton was dominated by John Wheeler. Although he was highly overcommitted and frequently out of town or otherwise unavailable, he was a wonderful advisor in that he generally treated his graduate students as his superiors, he was very supportive and encouraging, he was truly open to all ideas, and he had boundless energy and enthusiasm. At the time, however, I was not particularly enamored by his style of doing physics. To me, some of his grand ideas seemed to be over-reaches, and I preferred working on problems that were sufficiently well posed to have a well defined mathematical resolution.

One of the problems of great interest to me and others at Princeton at the time was the idea that, in Wheeler's phrase, ``black holes have no hair"---that a black hole, once formed or perturbed, will approach a stationary final state that is uniquely determined by its mass, angular momentum, and electric charge. It is here that Bekenstein made his first major contribution to general relativity: He showed that black holes cannot have ``scalar hair"; i.e., there cannot exist a nonvanishing stationary Klein-Gordon scalar field around a stationary black hole \cite{bek1}. Given the general difficulty of proving black hole uniqueness results, I was quite amazed by how simple and elegant Bekenstein's proof was. He realized that there was no need to use Einstein's equation; the Klein-Gordon equation alone in an arbitrary static or stationary-axisymmetric black hole background was sufficient to rule out time independent scalar field solutions that are well behaved at infinity and at the horizon.

The ``no hair" ideas had a very interesting implication: If one formed one black hole by the collapse of neutrons and another by the collapse of photons, the final state black holes would be exactly the same as long as they had the same total mass and angular momentum. Thus, the information about the original number of baryons present would be lost. This implied that the law of baryon conservation and, similarly, lepton conservation---which, at the time, were believed to be exact---would essentially become meaningless when black holes are present. Although no local violation of these laws would occur, one would have no way, in principle, of verifying that they were satisfied. Wheeler's term for this phenomenon was that in the presence of black holes, the laws of baryon and lepton conservation are ``transcended."

Wheeler was quite comfortable with the idea of transcendence of the laws of baryon and lepton conservation. However, it is quite remarkable (particularly in retrospect) that he was not comfortable with a similar transcendence of the second law of thermodynamics: If one takes a ``cup of tea"---the example Wheeler used---and throws it into a black hole, the entropy of that cup of tea will be lost, and total entropy of the universe will apparently decrease. One could maintain that the entropy is ``really still there" inside the black hole---just as one could maintain that the baryons are ``still there"---but, at best, the second law of thermodynamics would become observationally meaningless. My own view at the time was that the second law of thermodynamics is a statistical law, not a fundamental law, so its ``transcendence'' would be more palatable than the transcendence of an apparently fundamental law like baryon conservation. Thus, I was quite comfortable with the transcendence of the second law of thermodynamics. But Wheeler did not feel this way.

Wheeler suggested to Bekenstein that he look into ways that the second law might be rescued. I don't know whether it was Bekenstein or Wheeler who first thought of considering black hole area as possibly relevant to this issue, but it was a natural idea to consider at the time in this context: In 1970, Christodoulou had shown that one could assign an ``irreducible mass" to a Kerr black hole that could not decrease in any process involving dropping particle matter into the black hole \cite{chr}. Very shortly thereafter, Hawking proved his famous ``area theorem'' that showed, in complete generality, that the surface area of the event horizon of a black hole cannot decrease with time \cite{haw1}. Christodoulou's ``irreducible mass'' of a Kerr black hole was just its area. A non-decreasing quantity like black hole area would be a good place to look for something that might rescue the second law.

However, at the time, I felt that this was an utterly ridiculous project to work on. First, as already mentioned, I was not troubled by the apparent transcendence of the second law of thermodynamics. Second, the analogy between the second law and the area theorem seemed extremely artificial; it seemed quite unnatural to me to try to marry a statistical law with a mathematical theorem. But, most importantly, in the absence of a fully developed quantum theory of gravity, what could one possibly show and/or how could one possibly argue for the validity of any highly speculative ideas on black hole entropy that one might propose? I have a distinct memory of leaving a group meeting of Wheeler and all of his students shortly after Bekenstein had started working on this project, and seeing Bekenstein and Wheeler stay behind to talk further about the second law ideas. The thought that flashed through my mind at the time was ``Boy, am I glad that I have some good problems to work on!''

In December, 1971---when Bekenstein was deeply involved in developing his ideas on area representing entropy and on the generalized second law---Bob Geroch visited Princeton to give a colloquium. This was the first time I had met Geroch, and I was immediately taken with his style of doing physics and his infinite supply of fascinating ideas. The colloquium he gave remains one of the best I have ever heard. It was a very wide ranging talk that touched on many different aspects of general relativity. Part of the talk discussed black holes, and he spent perhaps 10 minutes making the point that one could use a black hole to run a Carnot cycle with 100\% efficiency. One could do this by lowering a box of high entropy matter all the way to the horizon of the black hole before dropping it in. All of the ``heat'' of the matter could thereby be converted to ``work" in the laboratory from which one did the lowering. Thus, Geroch pointed out that, in this well defined sense, black holes are systems at absolute zero temperature. I recall talking briefly with Bekenstein immediately after the colloquium. I was in high spirits about what a great talk it was. Bekenstein, of course, also thought it was a great talk, but he did not seem in as high spirits and, indeed, seemed a bit worried/concerned.

In retrospect, it seems clear that Bekenstein must have immediately realized that assigning an absolute zero physical temperature to a black hole would lead to severe consistency problems with black hole thermodynamics. In particular, Geroch's suggestion of lowering a box of matter containing entropy all the way to the horizon of a black hole could certainly be used to violate any proposal for a generalized second law, since, in this process, entropy would be lost, but the black hole would end up in the same state in which it began. Hence, whatever entropy, $S_{\rm BH}$, one might try to assign to the black hole, if one performed this process carefully enough, $S + S_{\rm BH}$ would necessarily decrease. Thus, Bekenstein had his work cut out for him if he was to salvage his ideas.

Bekenstein did salvage his ideas, using his considerable physical insight and understanding of what effects might be relevant to the issue. First, he argued that black hole entropy should be linearly proportional to area $A$ (rather than being a nonlinear monotonic function of $A$). On dimensional grounds, the entropy then had to be {\em inversely} proportional to $\hbar$---a rather unusual idea! This meant that the entropy of a macroscopic black hole would be absolutely enormous compared with the entropy of typical ordinary matter of comparable size and energy. But this meant, in turn, that to violate the generalized second law by lowering matter to towards a black hole, one had to bring the matter {\em extremely} close to the horizon before dropping it in---so close that it was not clear that this would be possible to do this for physical matter. This basic idea to use limitations on the properties of physical matter to rescue the generalized second law was clearly present in Bekenstein's original papers \cite{bek2}, \cite{bek3}. It became more formalized in his 1981 paper as the ``Bekenstein bound,'' $S/E \leq 2 \pi R$, on the ratio of entropy, $S$, to energy, $E$, for matter confined to a region of size $R$ \cite{bek4}. 

Nevertheless, black hole thermodynamics really is inconsistent if one assigns a finite entropy but a vanishing temperature to a black hole. This fact was sharpened in the remarkable paper of Bardeen, Carter, and Hawking, which showed that black holes obey precise analogs of the laws of thermodynamics \cite{bch}. The fact that black holes obey such laws was, in some sense, supportive of Bekenstein's thermodynamic ideas. However, by obtaining a precise first law of black hole mechanics, it was clear that if one wishes to assign a physical entropy to a black hole proportional to its area, $A$, then one must also must also assign to it a nonzero temperature proportional to its surface gravity, $\kappa$. But the fact that the physical temperature of a black hole vanishes then can be used to obtain contradictions with the generalized second law. In particular, suppose the black hole is placed in a thermal bath of radiation at a lower temperature than the assigned temperature of the black hole. Although the black hole may be smaller than typical wavelengths in this bath, it nevertheless, will still absorb some radiation. The generalized second law will then be violated, as heat will flow from the cooler to the hotter body. Indeed, Bardeen, Carter, and Hawking explicitly noted (using a Geroch-type box lowering argument rather than this thermal bath argument) that the horizon area of a black hole could not represent its physical entropy, and that the second law of thermodynamics is ``transcended'' in the presence of a black hole.

Sometime around 1973, I recall Bekenstein also telling me that his work had independently come under attack by a physicist for, I believe, the information theoretic ideas that he had used to assign an entropy to a black hole and to justify its thermodynamics. (I do not recall the name of this physicist or his specific criticism---he was not a general relativist and I had not heard his name before or since.) So, it certainly seemed that the main ideas that Bekenstein had worked on to date were not panning out very well! As already indicated above, I had not thought that assigning to a black hole a physical entropy proportional to its area made much sense in the first place---although I thought that Bekenstein had done a remarkably good job in advancing and defending this idea---and, as far as I was concerned, the paper by Bardeen, Carter, and Hawking provided a definitive refutation of the idea that the area of a black hole could represent its physical entropy.

Then, a miracle occurred! In 1974, Hawking calculated particle creation effects for a body that collapses to a black hole, and he made the amazing discovery that a distant observer will see a steady, thermal distribution of particles emerge at a temperature $T= \kappa/2 \pi$ \cite{haw2}. So, a black hole truly has a nonzero {\em physical} temperature proportional to its surface gravity! Black hole thermodynamics now appeared to be entirely consistent. In particular, if one placed a black hole in a radiation bath of temperature $T_{\rm bath} < \kappa/2 \pi$ the black hole radiation would dominate over absorption, and the generalized second law would hold. The entropy $S_{\rm BH} = A/4$ could now be interpreted as the {\em physical} entropy of the black hole---with the unknown constant in Bekenstein's original proposal now fixed by the value of the Hawking temperature. Bekenstein was right!

However, there remained a problem. Although the Hawking radiation by a black hole was sufficient to directly rule out any contradiction with the generalized second law if one immersed a black hole in a thermal bath of radiation, the Geroch-type box lowering argument still posed a potential difficulty for the generalized second law: As Geroch had originally argued, if one could lower a box arbitrarily close to the horizon, one should still be able get rid of the entropy in the box without increasing the black hole area. Here, it would seem, Hawking radiation or other quantum effects would not help, since one could perform this process for an arbitrarily large black hole, whose Hawking temperature is arbitrarily small. Thus, it appeared that one would have to rely on Bekenstein's proposed bound, $S/E \leq 2 \pi R$, to prevent one from lowering the box close enough to the horizon to violate the generalized second law. However, I was unhappy with this possible resolution for the following two reasons: (i) It did not appear that this bound was sufficient to avoid a violation of the generalized second law; if one used, e.g., a rectangular box, it would be necessary for the quantity ``$R$'' in the bound to be the {\em shortest} dimension of the box, whereas the arguments in favor of the bound took $R$ to be the {\em largest} dimension. (ii) It seemed to me that the consistency of black hole thermodynamics should not hinge on some property of matter that would not otherwise be needed for the consistency of thermodynamics.

Bill Unruh shared these concerns. Remarkably, we independently and simultaneously came up with a resolution in 1981---one week before I visited him in Vancouver: Although quantum effects such as Hawking radiation are negligible for freely falling observers near a large black hole, they are not negligible for quasi-stationary bodies near the horizon, such as a box of matter being slowly lowered towards the horizon. Such bodies undergo a huge acceleration and therefore feel the effects of ``acceleration radiation,'' which produces a buoyancy force. As we showed, as a consequence of this buoyancy force, the optimal place from which to drop a box of matter into the black hole is not the horizon but rather the ``floating point'' of the box. When the box is dropped from its floating point, the area increase of the black hole is exactly sufficient to compensate for the loss of the matter entropy, independently of whether or not any entropy bounds are satisfied by the matter \cite{uw1}.

 We immediately wrote to Bekenstein to tell him that the consistency of black hole thermodynamics did not rely on their being an entropy bound on ordinary matter. Bekenstein agreed with the validity of the buoyancy effect that we had found, but he felt that our approximations to describe it were not adequate and that some entropy bound on matter---although perhaps a weaker bound than the one he had previously proposed---would still be needed for the validity of the generalized second law. There followed a dueling series of papers \cite{bek5}-\cite{bek7} where Bekenstein gave examples where our approximations would not be adequate and argued that buoyancy would not be sufficient for the validity of the generalized second law, whereas we/I countered by showing that when the full calculation is done, buoyancy is sufficient. We never reached agreement on this issue.

Bekenstein and I remained friends throughout the years, but I saw him very infrequently after he moved to Israel in 1974. I was very pleased to come to Jerusalem in 2012 to celebrate the 40th anniversary of his original proposal that black hole area should represent its entropy. Although by then Bekenstein had won the Wolf Prize and many other high honors, he was as unassuming as he had been as a graduate student. He also was much more eager to discuss his new ideas than revel in past glories. I was also very pleased to be able to congratulate him in person at the American Physical Society meeting in 2015 upon his winning the Einstein Prize. It is tragic that he passed away only six months later.

\smallskip

\noindent
{\bf Acknowledgments}

This research was supported in part by NSF Grant PHY 15-05124 to the University of Chicago.


\begin{thebibliography}{99}

\bibitem{bek1}
J.D. Bekenstein, Phys.Rev. {\bf D5}, 1239 (1972).

\bibitem{chr}
D. Christodoulou, Phys. Rev. Lett. {\bf 25}, 1596 (1970).

\bibitem{haw1}
S.W. Hawking, Phys. Rev. Lett. {\bf 26}, 1344 (1971).

\bibitem{bek2}
J.D. Bekenstein, Phys.Rev. {\bf D7}, 2333 (1973).

\bibitem{bek3}
J.D. Bekenstein, Phys.Rev. {\bf D9},  3292 (1974).

\bibitem{bek4}
J.D. Bekenstein, Phys.Rev. {\bf D23}, 287  (1981).

\bibitem{bch}
J.M. Bardeen, B. Carter, and S.W. Hawking, Comm. Math. Phys. {bf 31}, 161 (1973).

\bibitem{haw2}
S.W. Hawking, Commun. Math. Phys. {\bf 43}, 199 (1975).

\bibitem{uw1}
W.G. Unruh and R.M. Wald, Phys. Rev. {\bf D25}, 942 (1982).

\bibitem{bek5}
J.D. Bekenstein, Phys.Rev. {\bf D27}, 2262 (1983).

\bibitem{uw2}
W.G. Unruh and R.M. Wald, Phys. Rev. {\bf D27}, 2271 (1983).

\bibitem{bek6}
J.D. Bekenstein, Phys. Rev. {\bf D49}, 1912 (1994).

\bibitem{pw}
M.A. Pelath and R.M. Wald, Phys. Rev. {\bf D60}, 104009 (1999) 

\bibitem{bek7}
J.D. Bekenstein, Phys.Rev. {\bf D60}, 124010 (1999).


\end{thebibliography}
\end{document}